\documentclass[runningheads]{llncs}
\usepackage{graphicx}
\usepackage{hyperref}
\usepackage{amsmath}
\usepackage{multirow}
\usepackage{tabularx}
\usepackage{float}
\usepackage{makecell}

\begin{document}

\title{Towards Optimal Patch Size in Vision Transformers for Tumor Segmentation}
\titlerunning{Optimal ViT Patch Size for Tumor Segmentation}
%
\author{Ramtin Mojtahedi\inst{1}\orcidID{0000-0002-3953-3256}, Mohammad Hamghalam\inst{1,2}\orcidID{0000-0003-2543-0712}, Richard K. G. Do\inst{3}\orcidID{0000-0002-6554-0310}, and Amber L. Simpson\inst{1,4}\orcidID{0000-0002-4387-8417}}
\authorrunning{R. Mojtahedi et al.}

%
\institute{School of Computing, Queen’s University, Kingston, ON, Canada
\email{amber.simpson@queensu.ca}\\
\and Department of Electrical Engineering, Qazvin Branch, Islamic Azad University, Qazvin, Iran
\and Department of Radiology, Memorial Sloan Kettering Cancer Center, New York, NY, USA
\and Department of Biomedical and Molecular Sciences, Queen’s University, Kingston,
ON, Canada}

\maketitle              
\thispagestyle{empty}
\begin{abstract}
Detection of tumors in metastatic colorectal cancer (mCRC) plays an essential role in the early diagnosis and treatment of liver cancer. Deep learning models backboned by fully convolutional neural networks (FCNNs) have become the dominant model for segmenting 3D computerized tomography (CT) scans. However, since their convolution layers suffer from limited kernel size, they are not able to capture long-range dependencies and global context. To tackle this restriction, vision transformers have been introduced to solve FCNN’s locality of receptive fields. Although transformers can capture long-range features, their segmentation performance decreases with various tumor sizes due to the model sensitivity to the input patch size. While finding an optimal patch size improves the performance of vision transformer-based models on segmentation tasks, it is a time-consuming and challenging procedure. This paper proposes a technique to select the vision transformer’s optimal input multi-resolution image patch size based on the average volume size of metastasis lesions. We further validated our suggested framework using a transfer-learning technique, demonstrating that the highest Dice similarity coefficient (DSC) performance was obtained by pre-training on training data with a larger tumour volume using the suggested ideal patch size and then training with a smaller one. We experimentally evaluate this idea through pre-training our model on a multi-resolution public dataset. Our model showed consistent and improved results when applied to our private multi-resolution mCRC dataset with a smaller average tumor volume. This study lays the groundwork for optimizing semantic segmentation of small objects using vision transformers. The implementation source code is available at: \url{https://github.com/Ramtin-Mojtahedi/OVTPS}.

\end{abstract}
\section{Introduction}
Colorectal cancer is the third most common cancer diagnosed in the United States, with 100,000 new cases and 50,000 deaths expected in 2022~\cite{ref_url1}. The survival rate of these patients is over 90\%~\cite{ref_url2}. However, up to 70\% of them will develop liver metastasis~\cite{ref_url3}, with a roughly 5-year survival rate of 11\%~\cite{ref_article1}. The segmentation of metastatic colorectal cancer (mCRC) liver tumours on computed tomography (CT) images is essential for evaluating tumour response to chemotherapy and surgical planning~\cite{ref_article2}, especially for detecting small metastasis tumor volumes in the liver tissue. To achieve this objective, it is imperative to build and develop a reliable and automated machine-learning (ML) model.

Convolutional neural networks (CNNs)-based \cite{ref_CNN1,ref_CNN2,ref_CNN3,ref_CNN4,ref_CNN5}  and vision transformers (ViTs)-based\cite{ref_TR1} architectures are the major machine learning segmentation approaches. Since the introduction of the pioneering U-shaped encoder-decoder architecture, dubbed U-Net \cite{ref_article3}, CNN-based architectures have achieved state-of-the-art performance on a variety of medical image segmentation tasks \cite{ref_article4}. The U-Net is a densely supervised encoder-decoder network where the encoder and decoder sub-networks are connected by densely supervised skip pathways. Adapting the U-Net to new challenges entails a range of design, preprocessing, training, and assessment strategies for the network. These hyperparameters are interconnected and have a substantial effect on the outcome. The nnU-Net framework was developed by Isensee et al. \cite{ref_article5} to address these limitations. Based on 2D and 3D vanilla U-Nets, they suggested nnU-Net as a robust and self-adaptive architecture.

Despite the effectiveness of fully convolutional networks, these networks have a drawback in learning global context and long-range spatial relationships due to their confined kernel size and receptive fields. To tackle this limitation, Dosovitsky et al. \cite{ref_article6} proposed using transformers in computer vision tasks, called ViTs, resulting from their successful performance in the language domain, their ability to capture long-range dependencies, and their self-attention mechanisms. Compared to state-of-the-art convolutional networks, ViT-based models achieve significant outcomes while using fewer computing resources for the training phase. By integrating an additional control mechanism in the self-attention module, a gated axial-attention model was presented by Valanarasu et al., \cite{ref_article7}, extending the previous transformer-based architectures. A novel ViT-based on a hierarchical structure was introduced by Liu et al. \cite{ref_article8} to represent the image features through shifted windows. Their proposed structure improved performance as self-attention processing is limited to non-overlapping local windows, but cross-window connections are still allowed. Hatamizadeh et al. \cite{ref_article9} proposed UNEt TRansformers (UNETR) to capture global multi-scale information. This unique U-Net-based architecture employs a transformer as the encoder to learn sequence representations of the input volume. The extracted features from the transformer encoder are integrated with the CNN-based decoder through skip connections to predict the segmentation outputs.

\raggedbottom Although the UNETR achieved state-of-the-art performance in 3D volumetric segmentation, it used an isotropic network topology with fixed-size feature resolution and an inflexible embedding size. Therefore, UNETR could not describe context at various sizes or assign computations at different resolutions. While their proposed network could be performant for segmenting large-sized objects such as the liver, it did not show the same level of performance in segmenting small objects such as small liver tumors.

Our goal in this paper was to detect and segment colorectal liver metastases on abdominal CT scans. The contribution of this work is twofold: First, we introduce a framework to find an optimal patch size for the vision transformer models to improve ViT-based structures in segmenting small objects. Specifically, based on the liver lesion volume, we designed a framework for ViT patch size, using the UNETR architecture as the backbone of our experiments to achieve higher segmentation performance. We also validated our proposed framework in a transfer-learning approach and showed that pre-training on training data that has a larger tumor volume using the proposed optimal patch size and then training with a smaller one achieved the best Dice similarity coefficient (DSC) performance. Second, we show that our pipeline outperforms the UNETR baseline ViT-based model in terms of DSC for segmenting liver metastasis and validates our results on LiTS and mCRC datasets.


\section{Method}
In the following subsections, the structure of the ViT-based model, our procedure to select optimal patch size, and a novel training technique to improve performance on segmenting of small tumors are elaborated.
\subsection{ViT-based Model Structure} \textbf{Transformer Patch}. In the ViT transformer framework, the input image is split into patches, and a series of linear embeddings of these patches is passed to the vanilla transformer~\cite{ref_article6}. These image patches are processed and considered similar to tokens (words) in natural language processing. Specifically, transformers operate on a 1D sequence of input embeddings. Similarly, the given 3D input images are mapped to 1D embedding vectors in our pipeline. In the utilized framework, the 3D CT volumes are provided with an input size of \textit{(H, W, L)} are the input image dimensions. The transformer patches are represented as \textit{M}. Accordingly, flattened uniform sequences are being created with the size of \textit{$N=(H\times$W$\times $L$)/(M\times$M$\times$M$)$} using non-overlapping patches that are shown with \textit{$y_v\in{R^{N\times M^3}}$}.

To maintain the retrieved patches’ spatial information, 1D learnable positional embeddings ($\textbf{E}_{pos}\in R^{N \times P}$) 
are added to the projected patch embeddings with dimensional embedding size. This process is shown in Eq. \eqref{Eq1}. 

\begin{equation}\label{Eq1}
\mathbf{z}_0 = [y_v^1\mathbf{E};y_v^2\mathbf{E};...;y_v^N\mathbf{E}]+\mathbf{E}_{pos}
\end{equation}
where $\textbf{E}\in R^{M^3\times P}$is flattened uniform non-overlapping patches embedding [13]. The transformer encoder consists of alternating layers of multi-headed self-attention (MSA) and multilayer perceptron (MLP) blocks. As proposed in [9], these blocks can be shown as Eq. \eqref{Eq2} and Eq. \eqref{Eq3}.


\begin{equation}\label{Eq2}
\normalfont{
\acute{\mathbf{z}_j} = \textbf{MSA}(\textbf{Lnorm}(\mathbf{z}_{j-1}))+\mathbf{z}_{j-1}, j\in[1,...,T]
}
\end{equation}

\noindent 
\begin{equation}\label{Eq3}
\acute{\mathbf{z}_j} = \textnormal{\textbf{MLP}}(\textnormal{\textbf{Lnorm}}(\acute{\mathbf{z}_j}))+\acute{\mathbf{z}_j}, j\in[1,...,T]
\end{equation}

\begin{figure}[t!]
\begin{center}
\includegraphics[width=0.9\textwidth]{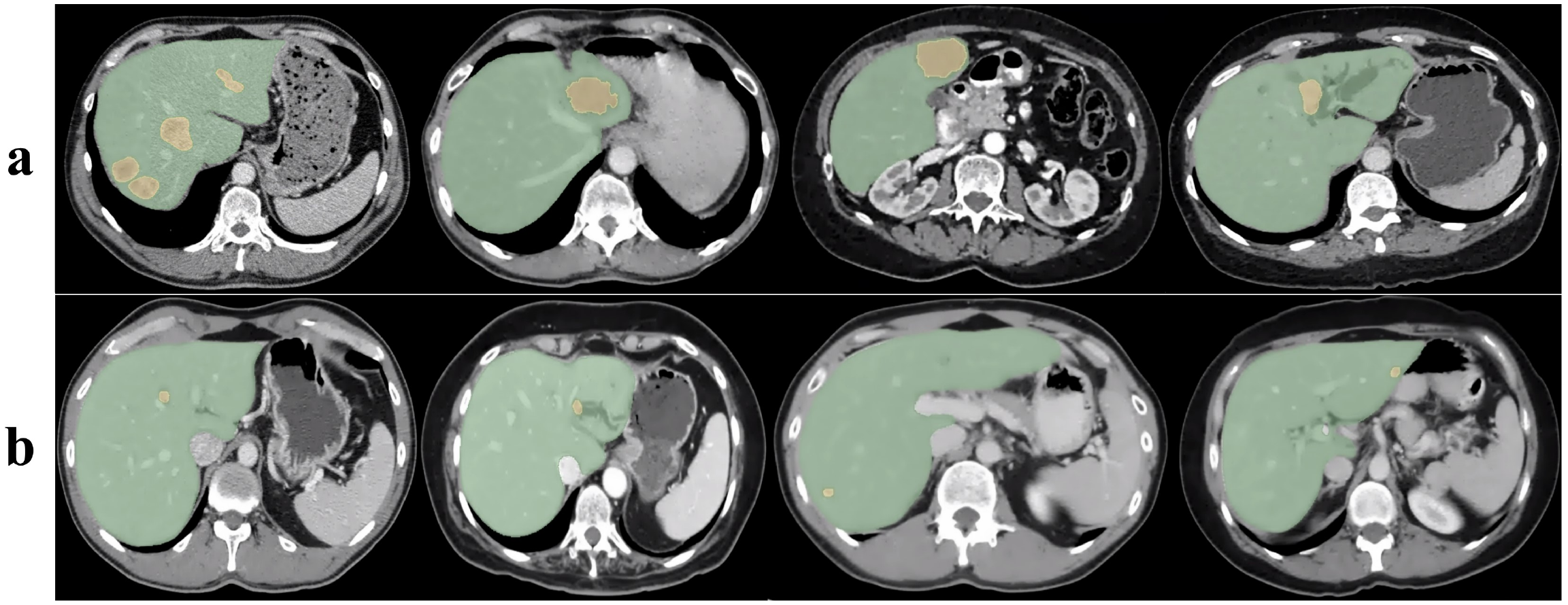}
\caption{Comparing tumor sizes in mCRC and LiTS sample data where (a) is the LiTS dataset with larger tumor volume size, and (b) is the mCRC dataset (Green: liver, Yellow: tumor).} \label{fig1}
\end{center}
\vspace{-1em}
\end{figure}

where \textit{T} is the number of transformer layers, \textit{Lnorm} function denotes layer normalization, and the MLP block consists of two linear layers with GELU activation functions. There are parallel self-attention (SA) heads in the MSA sublayer, and attention weights are calculated as (4). The SA block uses standard \textbf{qkv} self-attention, which uses query (\textbf{q}) and the sequence’s associated key (\textbf{k}) and input sequence value (\textbf{v}) representations. Eq. \eqref{Eq4} is also included  where $K_h=K/n$ is the scaling factor and the outputs of the MSA are achieved as Eq. \eqref{Eq5} using MSA weights.

\begin{equation}\label{Eq4}
\textnormal{Attention}(\textbf{q,k,v}) = \textnormal{Softmax}(\frac{\textbf{qK}^T}{\sqrt{K_h}})\textbf{v}
\end{equation}

\begin{equation}\label{Eq5}
[\textbf{Attention}_1(z);...;\textbf{Attention}_n(z)]\textbf{{W}}_{MSA}
\end{equation}
\subsubsection{Loss Function.}\normalfont{The loss function employed is a mix of soft Dice loss and cross-entropy loss, which could be calculated in a voxel-by-voxel approach, as shown in Eq. \eqref{Eq6}. 

\begin{equation}\label{Eq6}
\begin{split}
Loss(G,O)=1-\frac{2}{C}\sum_{r=1}^C\frac{\Sigma_{x=1}^A G_{x,r}O_{x,r}}{\Sigma_{x=1}^C G_{x,r}^2+\Sigma_{K=1}^U O_{x,r}^2}-\frac{1}{A}\sum_{x=1}^A\sum_{r=1}^C G_{x,r}log(O_{x,r})
\end{split}
\end{equation}
where \textit{A} represents the voxel’s number; \textit{C} denotes the number of classes. The probability output and one-hot encoded ground truth for class \textit{r} at voxel \textit{x} are represented by \textit{$O_{x,r}$} and \textit{$G_{x,r}$}, respectively~\cite{ref_article10}}.

UNETR uses a contracting-expanding pattern with a stack of transformers, ViT, as the encoder, and skip connections to the decoder. It considers the patch size as a hyperparameter. In this sense, choosing an optimal patch size (\textit{$M^{*}$}) is critical due to its impact on the features’ receptive field. This is because patches are reshaped into a tensor with the size of $\frac{H}{M^{*}}\times\frac{W}{M^{*}}\times\frac{L}{M^{*}}\times P$, where \textit{P} is the transformer’s embedding size. To assess the impact of patch size, we did our experiments on two clinical datasets, LiTS, and our private mCRC. LiTS has a larger tumors volume size than the mCRC dataset. Samples of these two datasets are shown in Fig.~\ref{fig1}.

\subsection{Choosing Optimal Patch size}
\begin{figure}[t!]
\includegraphics[width=\textwidth]{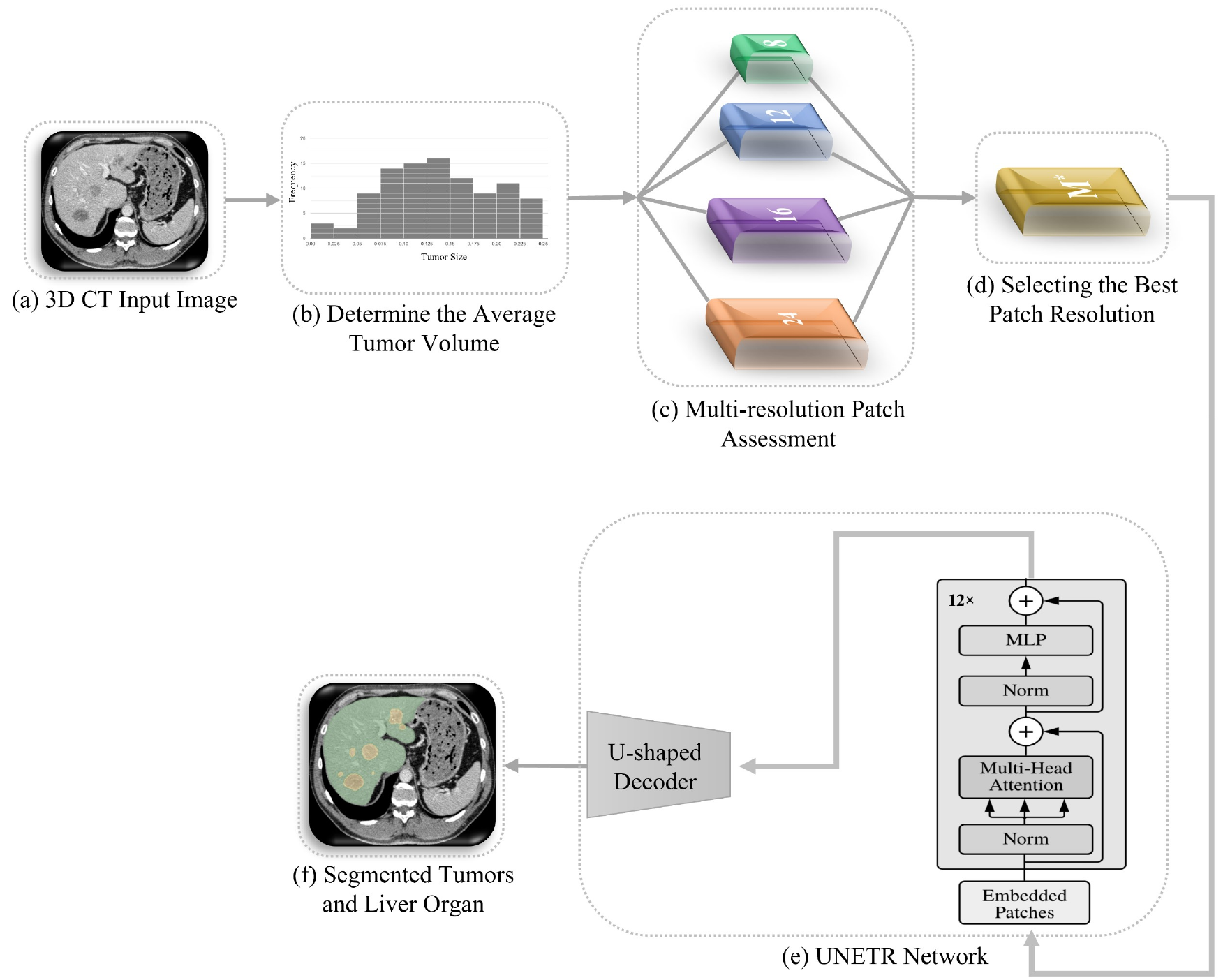}
\caption{The model’s pipeline is shown for the proposed framework. In step (a), the framework receives the raw 3D CT images of the abdomen, consisting of the liver and its primary and secondary tumors. Then, the average tumor volume is computed through histogram analysis in step (b). Through assessment of the averaged volume of tumors and the three determined patch sizes in step (c), \textit{$M\in[8, 12, 16, 24]$} the optimal patch size is selected in step (d). In step (e), we train the model and segment it in our backboned UNETR network to achieve segmented tumors and liver organs in step (f).} \label{fig2}
\end{figure}

The proposed framework tries to find the best patch size for our tumor segmentation. As shown in Fig.~\ref{fig2}, the average volume of the tumors is first computed on the training input. Then, the optimal patch size is determined based on a mathematical relationship between the average volume size of tumors and the performance. The experimented patch sizes must be a factor of the input image dimensions, $256\times256\times96$, and were selected based on our computational resources and with respect to the sizes proposed in~\cite{ref_article6}, \textit{$M\in[8, 12, 16, 24]$}. This ensures that the model performs best when segmenting small objects such as tumors.

The average volume size of tumors in LiTS datasets was reported as 17.56 $cm^{3}$~\cite{ref_article11}. Through histogram analysis on our private mCRC dataset, the average tumor volume was achieved as 10.42 $cm^{3}$. Empirically, the relationship between optimal patch size (\textit{$M^{*}$}) and the average volume size of the tumors attained as Eq. \eqref{Eq7}. 

\begin{equation}\label{Eq7}
M^{*} = \underset{M\in [8, 12, 16, 24]}{\mathrm{Argmin}} 
(|\sqrt[3]{V\times S} - M|)
\end{equation}
where \textit{$M$} is the patch size; \textit{$S$} is the voxel spacing, and \textit{$V$} is the average volume size of tumors for LiTS and mCRC datasets. The optimal patch is achieved by finding the patch that makes the absolute differentiation of cube root multiplication of voxel spacing and average tumors volume size with patch size be at a minimum. 

\subsection{Pre-training Technique to Improve Segmentation of Small Tumors}
To increase the segmentation performance of ViT-based structures on small lesions, we suggested pre-training on a dataset with a large tumor volume (LiTS data) using the optimal patch size achieved by the proposed framework and subsequently training on a smaller one (mCRC dataset) to achieve the best DSC performance. This idea was experimentally tested, as shown in the next section, and could increase the DSC significantly compared to when the dataset with a small tumor volume was trained by scratch.

\section{Experimental Results}
\subsection{Datasets}
We conducted our experiments on two multi-resolution 3D abdominal CT liver datasets, including training with large tumor volumes (LiTS) and later smaller ones (mCRC). For the former, we used the LiTS, which consisted of 201 CT images with liver and liver tumors annotations: 131 for training and 70 for testing. The number of tumors detected in the scans varied between 0 and 75, exhibiting a half-normal distribution. The dataset was created to closely show real-world clinical data and contains a range of cancer types, including primary tumors such as hepatocellular carcinoma (HCC) and metastasis from colorectal, breast, and lung cancer. The collection includes scans with voxel spacings ranging from 0.56 mm to 1.0 mm in the axial plane and slice thicknesses ranging from 0.70 mm to 5.0 mm~\cite{ref_article11,ref_article12}. In the private data, we employed CT volume of the colorectal liver metastasis (mCRC)~\cite{ref_article13}. This dataset contained 198 CT scans of patients who underwent hepatic resection for CRLM between 2003 and 2007. The number of tumors in the scans ranged from 1 to 17. Voxel spacing in the scans in the dataset ranged from 0.61 mm to 0.98 mm in the axial plane, and slice thickness was 0.80 mm to 7.5 mm.  For data pre-processing, all image voxel spacing was normalized to the range [0–1]. In addition, all foreground images were resampled to a voxel spacing of $0.765\times0.765\times1.5 mm^3$, achieved by the median of the range of spacings and availability of computational resources. The data is also transformed using $90$ orientation, flipping, random rotations, and intensity shifting. 

\subsection{Implementation details}
\raggedbottom Table~\ref{tab1} illustrates the important model parameters and hyperparameters we employed to conduct the experiments on our datasets. The hyperparameters used for the ViT network were selected based on the ViT-base discussed in ~\cite{ref_article6}. We also ran the experiments with various input image sizes and discovered that $256\times256\times96$ produced the best results compatible with our computational resources. Implementations of experiments and code will also be available. For all experiments, the training and validation split considered as 80:20. 

\begin{table}
\caption{Summary of employed parameters and critical hyperparameters.}\label{tab1}
\begin{tabular}{|l|l|}
\hline
\bfseries{Parameter} &  \bfseries{Description of the Value/Method}\\
\hline
Input Image Size $[H\times W\times L]$ &  $[256\times256\times96]$\\
\hline
Optimizer &  Adam\\
\hline
Learning Rate & 0.0001\\
\hline
Weight-Decay & 1e-5\\
\hline
\makecell{ViT: [Layers, Hidden Size, MLP size, Heads,\\ Number of Parameters] } 
& [12, 768, 3072, 12, 86M]\\
\hline
Batch Size & 1\\
\hline
Computational Resource & NVIDIA A100 – 40GB\\
\hline
\end{tabular}
\vspace{-8mm}
\end{table}

\subsection{Liver and Lesion Segmentation Results}
\textbf{Segmentation Results for Optimal Patch size.} We calculated the optimal patch size, \textit{$M^{*}$}, based on eq. (7) for both datasets. We also compared our results with smaller and larger patch sizes than the calculated one to validate our proposed technique. As shown in Table~\ref{tab2}, the best performance results were achieved for the computed patch size of 16 and 12 for the LiTS and mCRC datasets, respectively. In addition to these experiments, we tested the DSC performance using a combination of both datasets, which did not outperform the following results. Moreover, as our main focus was to find the optimal patch size, we didn't provide results with respect to the CNN-based architectures, which inherently have different structures with no utilized transformer. 

\begin{table}[H]
\centering
\caption{Highest segmentation performance  results for the models built on LiTS and mCRC datasets using multiple patch sizes.}\label{tab2}
\begin{tabular}{|c|cl|c|cl|c|cllll|}
\hline
\textbf{Dataset}      & \multicolumn{2}{c|}{\textbf{Patch Size}}   & \textbf{Tumor DSC {[}\%{]}} & \multicolumn{2}{c|}{\textbf{Liver DSC {[}\%{]}}} & \textbf{Loss}           & \multicolumn{5}{c|}{\thead{\textbf{Training}\\ \textbf{Time} {\textbf{[}Min.{]}}}} \\ \hline
\multirow{4}{*}{LiTS} & \multicolumn{2}{c|}{M=8}                   & 48.62                       & \multicolumn{2}{c|}{81.3}                        & 0.2105                  & \multicolumn{5}{c|}{1883.93}                           \\ \cline{2-12} 
                      & \multicolumn{2}{c|}{M=12}                  & 51.19                       & \multicolumn{2}{c|}{87.37}                       & 0.2297                  & \multicolumn{5}{c|}{2464.83}                           \\ \cline{2-12} 
                      & \multicolumn{2}{c|}{\textbf{M*=16}}        & \textbf{53.08}              & \multicolumn{2}{c|}{\textbf{88.06}}              & \textbf{0.1805}         & \multicolumn{5}{c|}{\textbf{2811.70}}                  \\ \cline{2-12} 
                      & \multicolumn{2}{c|}{M=24}                  & 51.91                       & \multicolumn{2}{c|}{87.93}                       & 0.1717                  & \multicolumn{5}{c|}{4106.87}                           \\ \hline
\multirow{5}{*}{mCRC} & \multicolumn{2}{c|}{M=8}                   & 39.64                       & \multicolumn{2}{c|}{89.51}                        & 0.1893                  & \multicolumn{5}{c|}{3745.30}                           \\ \cline{2-12} 
                      & \multicolumn{2}{c|}{\textbf{M*=12}}                  & \textbf{41.44}                       & \multicolumn{2}{c|}{\textbf{92.35}}                       & \textbf{0.1020}                  & \multicolumn{5}{c|}{\textbf{2221.24}}                           \\ \cline{2-12} 
                      & \multicolumn{2}{c|}{M=16}        & 40.14              & \multicolumn{2}{c|}{87.77}              & 0.1060         & \multicolumn{5}{c|}{\t2566.82}                  \\ \cline{2-12} 
                      & \multicolumn{2}{c|}{M=24}                  & 38.82                       & \multicolumn{2}{c|}{87.85}                       & 0.2050                  & \multicolumn{5}{c|}{3758.87}                           \\ \hline
\end{tabular}
\end{table}

\noindent \textbf{Effectiveness of the Proposed Vision-based Model Training.} Table~\ref{tab3} indicates that employing LiTS pre-trained models significantly improves segmentation performance. The patch size of 16 showed the best outcomes, with a DSC of 44.94 percent for tumor segmentation. This indicates that training on a large tumor volume dataset successfully learns tumor representations that improve model performance on an mCRC dataset with small tumor volume mCRC. 

\begin{table}
\centering
\caption{Comparison of the highest segmentation performance (DSC(\%)) results using the pre-trained model on the dataset with larger tumor volumes (LiTS) to the dataset with smaller tumor volumes (mCRC).}\label{tab3}
\begin{tabular}{|c|cc|cc|ccll|}
\hline
\multirow{2}{*}{\textbf{Patch Size}} & \multicolumn{2}{c|}{\textbf{Pre-trained Model}}              & \multicolumn{2}{c|}{\textbf{Non Pre-trained Model}}          & \multicolumn{4}{c|}{\textbf{Improvement}}                                         \\ \cline{2-9} 
                                     & \multicolumn{1}{c|}{Tumor} & Liver& \multicolumn{1}{c|}{Tumor } & Liver & \multicolumn{1}{c|}{Tumor} & \multicolumn{3}{c|}{Liver} \\ \hline
8                                    & \multicolumn{1}{c|}{42.2}               & 93.97              & \multicolumn{1}{c|}{39.64}              & 89.51              & \multicolumn{1}{c|}{2.56}               & \multicolumn{3}{c|}{4.46}               \\ \hline
12                                   & \multicolumn{1}{c|}{44.46}              & 94.42              & \multicolumn{1}{c|}{41.44}              & 92.35              & \multicolumn{1}{c|}{3.02}               & \multicolumn{3}{c|}{2.07}               \\ \hline
\textbf{16 (M*)}                          & \multicolumn{1}{c|}{\textbf{44.94}}     & \textbf{94.61}     & \multicolumn{1}{c|}{\textbf{40.14}}     & \textbf{87.77}     & \multicolumn{1}{c|}{\textbf{4.8}}       & \multicolumn{3}{c|}{\textbf{6.84}}      \\ \hline
\end{tabular}
\end{table}

\begin{figure}[t!]
\includegraphics[width=\textwidth]{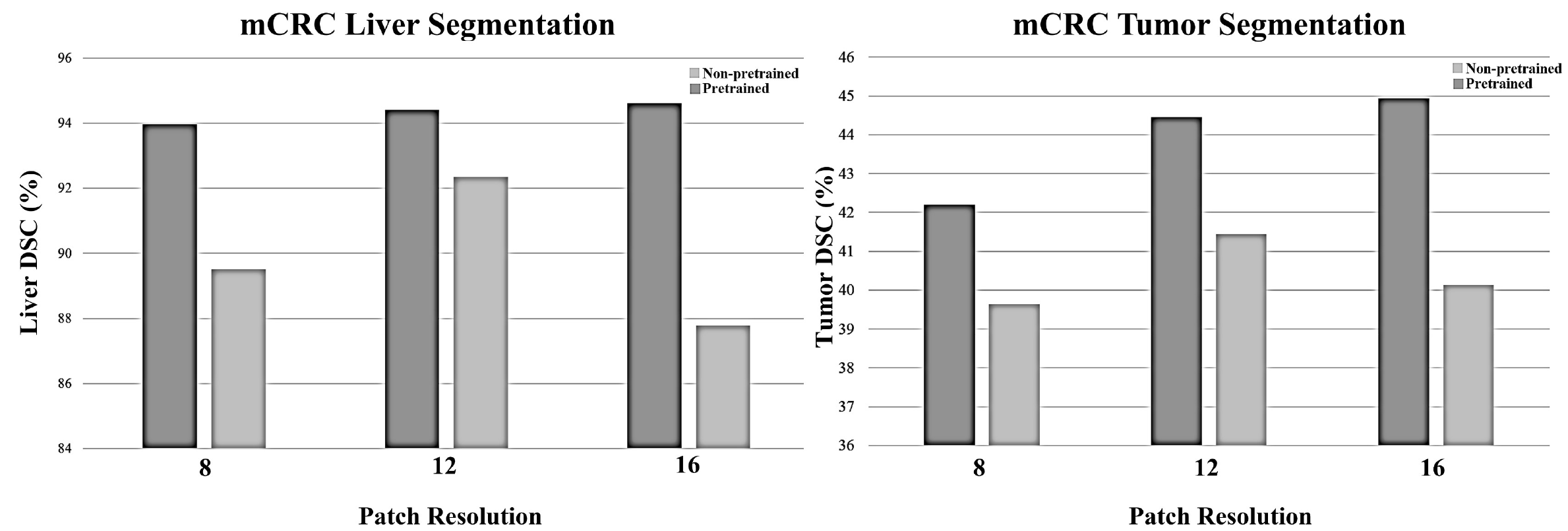}
\caption{The performance results for the tumor and liver organ segmentation tasks were obtained using LiTS pre-trained models and mCRC itself for training. All pre-trained models could improve performance, while the model with a patch size of 16 achieved the best results, improving the tumor segmentation performance by 4.8\%.} \label{fig3}
\end{figure}

Fig.~\ref{fig3} visually compares the segmentation performance between pre-trained models on the LiTS dataset with larger tumor volumes to the mCRC dataset with smaller tumor volumes both for tumor and liver organ.

\section{Discussion and Conclusion}
This paper proposed a novel framework to find an optimal patch size for semantic segmentation, particularly practical for small liver lesion segmentation. Based on the volume size of metastasis, we introduced a procedure to calculate patch size methodically in transformer-based segmentation models. In addition, the optimal patch size computed by the proposed method showed the best performance on large objects such as a liver organ. Furthermore, a significant part of the small tumor information was missed when we trained on a large patch size. However, when we pre-trained a model on our public dataset of LiTS with larger tumors, the model could learn tumors representations with higher performance. Consistent with our first novelty and in a transfer-learning approach, the pre-trained model demonstrated its most effective performance in learning representations and segmentation performance when it utilized the computed optimal patch size defined by (7), \textit{$M^{*}$}. The results of this study could be used for further development in vision transformer-based networks with multi-patch sizes. We also showed that our pipeline outperforms the ViT-based models in terms of DSC for segmenting liver metastasis tumors and validated our results on LiTS and mCRC datasets. 

\section{Acknowledgement}
This work was funded in part by National Institutes of Health R01CA233888.

%
%
%
%
%

\end{document}